\documentclass[aps,prl,final,twocolumn,letterpaper]{revtex4-1}

\usepackage{graphicx}   
\usepackage{import}                         
\usepackage{epstopdf}
\usepackage{amsmath} 
\usepackage{float} 
\usepackage{bm}
\usepackage{amssymb}
\usepackage{quotes}
\usepackage{indentfirst}
\usepackage{color}
\usepackage{transparent}
\usepackage{dcolumn}
\usepackage{braket}
\usepackage{multirow}
\usepackage{cancel} 
\usepackage{mdframed}
\usepackage{color}
\usepackage{bm}
\usepackage{dsfont}
\usepackage{slashed}
\usepackage{soul, color} 
\soulregister\ref{7}  
\soulregister\cite{7} 
\renewcommand{\st}[1]{}

\usepackage{xr}

\usepackage[version=4]{mhchem}

\makeatletter
\newcommand*{\addFileDependency}[1]{
  \typeout{(#1)}
  \@addtofilelist{#1}
  \IfFileExists{#1}{}{\typeout{No file #1.}}
}
\makeatother

\newcommand*{\myexternaldocument}[1]{%
    \externaldocument{#1}%
    \addFileDependency{#1.tex}%
    \addFileDependency{#1.aux}%
}

\myexternaldocument{si}

\usepackage{textcomp} 
\usepackage{xifthen}
\usepackage{xcolor}
\usepackage{etoolbox}
\newboolean{togglechanges} 
\usepackage{siunitx}
\usepackage{wrapfig}

\setboolean{togglechanges}{false}

\newcommand{\comment}[1]{\ifbool{togglechanges}
    {#1}  
    {\textcolor{blue}{#1}}}

\usepackage{bibentry}

\usepackage{graphicx}
\usepackage{dcolumn}
\usepackage{bm}
\setcitestyle{numbers,square,super}

\usepackage{textcomp} 
\begin{document}
\rmfamily

\title{Hybrid Nanophotonic Scintillators for Enhanced X-ray Absorption, Emission, and Time Resolution}
\author{Seou~Choi$^{1}$}
\email{seouc130@mit.edu}
\author{Sachin~Vaidya$^{1,2}$}
\author{Avner Shultzman$^{3}$}
\author{Charles~Roques-Carmes$^{1,4}$}
\author{Ido Kaminer$^{3}$}
\author{Marin~Solja\v{c}i\'{c}$^{1,2}$}
\affiliation{$^{1}$ Research Laboratory of Electronics, Massachusetts Institute of Technology, Cambridge, MA 02139, USA\looseness=-1}
\affiliation{$^{2}$ Department of Physics, Massachusetts Institute of Technology, Cambridge, MA 02139, USA\looseness=-1}
\affiliation{$^{3}$ Department of Electrical and Computer Engineering, Technion - Israel Institute of Technology, Haifa 32000, Israel\looseness=-1}
\affiliation{$^{4}$ E. L. Ginzton Laboratories, Stanford University, Stanford, CA 94305, USA\looseness=-1}

\clearpage 

\setlength{\parskip}{0em}
\vspace*{-2em}


\vspace{0.8cm}

\begin{abstract}

    Scintillators convert ionizing radiation into visible photons, enabling applications from cosmic ray detection to medical imaging. Two independent strategies for improving scintillator performance via nanoscale patterning have recently been demonstrated: engineering material properties to enhance absorption of ionizing radiation and integrating nanophotonic structures to enhance the spontaneous emission rate (``nanophotonic scintillators''). Here, we propose a nanophotonic scintillator that simultaneously enhances both the initial energy conversion and the spontaneous emission rate, by periodically stacking a fast-emitting scintillator and a visible-light-transparent material with strong X-ray attenuation (``stopping layer'') to form a one-dimensional (1D) photonic crystal (PhC) scintillator. Photoelectric absorption in the stopping layer increases the number of photoelectrons that deposit energy in neighboring scintillator layers and contribute to scintillation. At the same time, the spontaneous emission rate is enhanced by the nanophotonic structuring itself. We design a 1D PhC comprising an organic scintillator and indium tin oxide (ITO) as the stopping layer and numerically simulate the enhancement in scintillation yield and decay rate. The total detected light output is enhanced by up to a factor of 700 compared to a bulk organic scintillator of equal thickness. We further investigate a 1D PhC structure integrating inorganic and organic scintillators for time-of-flight positron emission tomography (TOF-PET): replacing the non-scintillating stopping layer with an inorganic scintillator further increases the light yield, and the coincidence time resolution (CTR) is enhanced up to 3.5 times compared to a bulk inorganic scintillator of equal thickness. Our work presents a unified approach to improve key scintillation parameters within a single nanophotonic structure.

\end{abstract}

\maketitle

\section*{Introduction}

Scintillators are essential components in high-energy particle detection~\cite{moses2002current}, medical imaging~\cite{van2002inorganic}, and homeland security~\cite{glodo2017new}, where they enable the detection of ionizing radiation such as X-rays or high-energy electrons. These materials absorb ionizing radiation and subsequently emit visible light through spontaneous emission. To maximize light output, scintillators must efficiently absorb the energy of incident high-energy particles. The scintillation decay time, which governs the maximum detection rate, is fundamentally determined by the radiative lifetime of the luminescent center. Achieving both high light output and short decay time remains a central challenge in scintillator development. Early approaches to enhancing scintillator performance have focused on engineering intrinsic material properties to optimize light output and decay time~\cite{chen2018all,wei2016sensitive,du2024efficient}. Recently, enhancing energy absorption using nanoscale heterostructures has also been explored~\cite{be2025heterostructure}. 

An alternative strategy is to integrate nanophotonic structures that modify the emission properties of the scintillator~\cite{kurman2020photonic}. Nanophotonic structures such as photonic crystals~\cite{knapitsch2014review} and plasmonic structures~\cite{ye2024nanoplasmonic,liu2017plasmonic} fabricated on the scintillator backplane can enhance the fraction of scintillating light escaping the scintillator (outcoupling efficiency). Advances in nanofabrication techniques enabled direct patterning of scintillating materials at the nanoscale~\cite{kurman2024purcell,li2024fabrication,martin2025large,roques2022framework} or incorporating nanoparticles inside bulk scintillators~\cite{makowski2025scaling}, improving spontaneous emission through the Purcell effect. 

Nevertheless, these two approaches have so far been investigated independently, due to the disparate length scales required for efficient scintillation: efficient X-ray absorption requires a scintillator thickness comparable to the X-ray attenuation length ($>$\SI{10}{\micro\meter} for hard X-rays), whereas nanophotonic structures operate at length scales comparable or smaller than the order of the visible light wavelength. A nanophotonic scintillator structure that facilitates efficient X-ray absorption would enable both rapid and high-yield scintillation.

Here, we present a nanophotonic scintillator that simultaneously enhances the energy conversion efficiency and reduces the decay time by leveraging both the photoelectric effect and Purcell effect. This could be achieved with a one-dimensional (1D) photonic crystal (PhC) formed by periodically stacking fast-emitting scintillator and a visible-light-transparent material with strong X-ray attenuation, which we refer to as a ``stopping layer''. We first demonstrate that nanoscale stopping layers substantially enhance the number of scintillating photons generated inside the structure. While the stopping layers efficiently absorb incident ionizing radiation via the photoelectric effect, interleaving them at nanoscale thickness with a bulk organic scintillator can maximize the number of photoelectrons that contribute to scintillation. Simultaneously, the nanoscale periodic placement of stopping and scintillator layers enhances spontaneous emission through the Purcell effect.

We first showcase our concept by combining an organic scintillator with indium tin oxide (ITO). ITO possesses a high refractive index and strong X-ray attenuation, which makes it a suitable material for the stopping layers. Numerical simulation results show that the optimized PhC structure yields up to a 700-fold increase in scintillation light output compared to a bulk organic scintillator with an equal thickness, together with a 1.4 times faster decay time through Purcell enhancement. We then extend our concept to a ``hybrid PhC scintillator'' in which an inorganic scintillator serves as the stopping layer. This hybrid structure combines the high scintillation yield of inorganic scintillators and fast decay times of the organic scintillator. We showcase that the hybrid PhC scintillators achieve 3.5 times enhancement in the coincidence time resolution (CTR) suitable for time-of-flight positron emission tomography (TOF-PET)~\cite{vandenberghe2016recent}.

\begin{figure}[t]
    \centering
    \includegraphics[scale = 1.1]{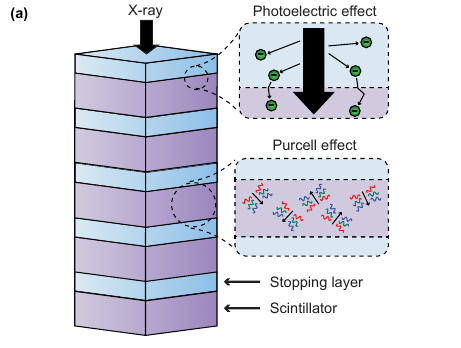}
    \vspace*{-4mm}
    \caption{\textbf{Photoelectric-enhanced nanophotonic scintillators.} When an X-ray hits the stopping layer (e.g., indium tin oxide (ITO)), it generates electrons through the photoelectric effect. High-energy photoelectrons that can escape the stopping layer enter the scintillator and generate visible photons. Scintillators with weak X-ray absorption such as organic scintillators can significantly enhance the scintillation light output with the aid of stopping layers. The thickness of the stopping layer and the scintillator layer are nanoscale, enabling faster scintillation via the Purcell enhancement.
}
    \label{fig:concept}
\end{figure}

\section*{Main} 
\subsection*{Concept of photoelectric-enhanced nanophotonic scintillators}

The scintillation process begins with the absorption of incident ionizing radiation, which generates high-energy electrons through the photoelectric effect or Compton scattering~\cite{knoll2010radiation}. These electrons undergo a series of inelastic scattering events until their energy is converted into visible photons. The resulting spontaneous emission is governed by the intrinsic radiative properties of the scintillator and by the Purcell effect, which describes the modification of the spontaneous emission rate due to the local optical environment~\cite{purcell1946resonance}.

Figure~\ref{fig:concept} describes how a nanophotonic scintillator structure can enhance the scintillation light output and the emission rate by increasing photoelectric absorption and the Purcell enhancement simultaneously. This is achieved in a 1D PhC scintillator comprising alternating layers of an organic scintillator and a material with high X-ray stopping power and high transparency in the visible spectrum (stopping layer).  The stopping layer efficiently converts the energy of incident ionizing radiation into photoelectrons, which can escape into the adjacent organic scintillator layers and provide energy to the luminescence centers. Rather than placing a single thick (bulk) stopping layer on top of the scintillator, we interleave subwavelength-thick stopping layers with a bulk organic scintillator such that photoelectrons can escape the stopping layer before losing their energy within it. As a result, this design significantly enhances the energy conversion efficiency and thereby increases the scintillation light output of organic scintillators.  

While both organic and inorganic scintillators can benefit from integration with stopping layers, inorganic scintillators already generate a large number of photoelectrons without stopping layers due to their high density and high effective atomic number. On the other hand, organic scintillators which often consist of light atoms cannot effectively attenuate X-rays through the photoelectric effect and rely primarily on Compton scattering. Compton scattering exhibits lower scattering cross sections than the photoelectric effect and deposits only a fraction of the incident photon energy into the electron, which results in a reduced scintillation output relative to inorganic scintillators. Therefore, we mainly focus on organic scintillators and investigate how their performance can be improved via photoelectric absorption and nanophotonic enhancement. 

\begin{figure*}
    \centering
    \includegraphics[scale = 1.0]{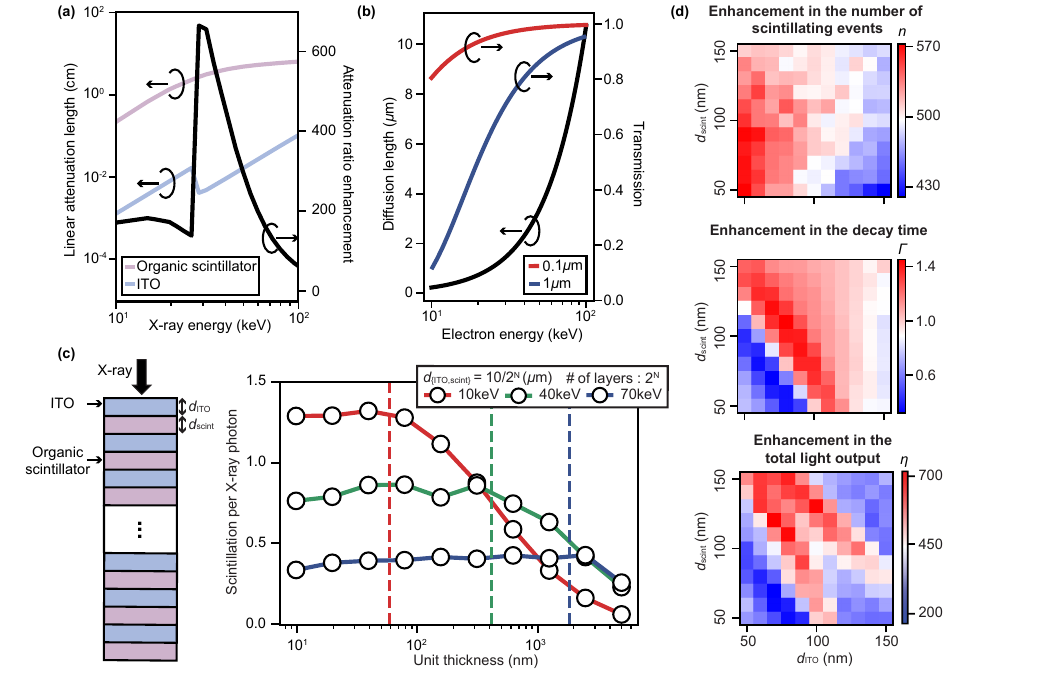}
    \vspace*{-4mm}
    \caption{\textbf{Scintillation performance enhancement in 1D photonic crystal (PhC) scintillators.}  (\textbf{a}) Linear attenuation length of indium tin oxide (ITO) and organic scintillator as a function of X-ray energy. (\textbf{b}) Diffusion length of electrons as a function of energy. (\textbf{c}) Number of scintillation events inside the 1D scintillator structure numerically calculated as a function of the thickness of unit layer under different incident X-ray energies. The total thickness is \SI{10}{\micro\meter}, and the thickness of each ITO and organic scintillator are set to be equal $d_{\textrm{scint}}=d_{\textrm{ITO}}=10/2^N$um, where $2^N$ is the number of layers ($N=1,2,3,...)$. Dashed lines in the right panel show the 1\% of the X-ray attenuation length of ITO at a given X-ray energy. (\textbf{d}) Enhancement of different scintillator parameters (number of scintillation events inside the structure $n$, decay time $\Gamma$, and total light output $\eta$) as function of $d_{\textrm{ITO}}$ and $d_{\textrm{scint}}$. The X-ray energy is fixed to \SI{30}{\kilo\eV} and PhC scintillators have 20 unit cells. Organic scintillators are modeled with EJ-296 from Eljen Technologies.}
    \label{fig:ITO_plastic}
\end{figure*}

\subsection*{1D photonic crystal scintillators consist of indium tin oxide and organic scintillator}

We now demonstrate how a subwavelength-thick stopping layer can enhance X-ray absorption and provide photoelectrons to the scintillator. Although various heavy metal oxides~\cite{lezal2001heavy} and halide crystals~\cite{wallach2025highly} possess strong X-ray attenuation and high transparency in the visible spectrum, we select indium tin oxide (ITO) as the stopping layer. ITO offers high photoelectric absorption due to its moderately high effective atomic number ($Z_{\textrm{eff}}=45.5$) and high density~\cite{murty1965effective}. In addition, ITO exhibits low optical absorption allowing the scintillating light to propagate through the entire structure, and a high refractive index in the visible spectrum, which increases the refractive index contrast and thereby strengthens nanophotonic control of spontaneous emission. The organic scintillator is modeled with EJ-296 plastic scintillator from Eljen Technologies. Figure~\ref{fig:ITO_plastic}(a) shows the X-ray attenuation length of ITO and the organic scintillator as a function of incident X-ray energy~\cite{jackson1981x}.  Due to the K-edge absorption~\cite{hubbell1996tables} near \SI{28}{\kilo \eV}, the X-ray attenuation length of ITO can be up to approximately 600 times shorter than that of the organic scintillator. After photoelectric absorption, the photoelectron carries the incident X-ray energy minus the binding energy of the atom. Fig.~\ref{fig:ITO_plastic}(b) shows the electron diffusion length~\cite{lukiyanov2009depth} inside ITO as a function of energy (left axis), along with the transmission probability of photoelectrons through ITO layers of two thicknesses (\SI{100}{\nano \meter} for red,  \SI{1}{\micro \meter} for blue). For an ITO layer thickness of \SI{100}{\nano \meter}, which is close to $\lambda/2n_{\textrm{ITO}}$ where $\lambda$ is the peak emission wavelength of the organic scintillator, more than 80\% of electrons escape the ITO. In contrast, the transmission significantly drops for \SI{1}{\micro \meter} thick ITO. This result highlights the importance of employing a subwavelength thick ITO layer to support efficient leakage of photoelectrons into the adjacent organic scintillator layers.

\begin{figure*}
    \centering
    \includegraphics[scale = 1.0]{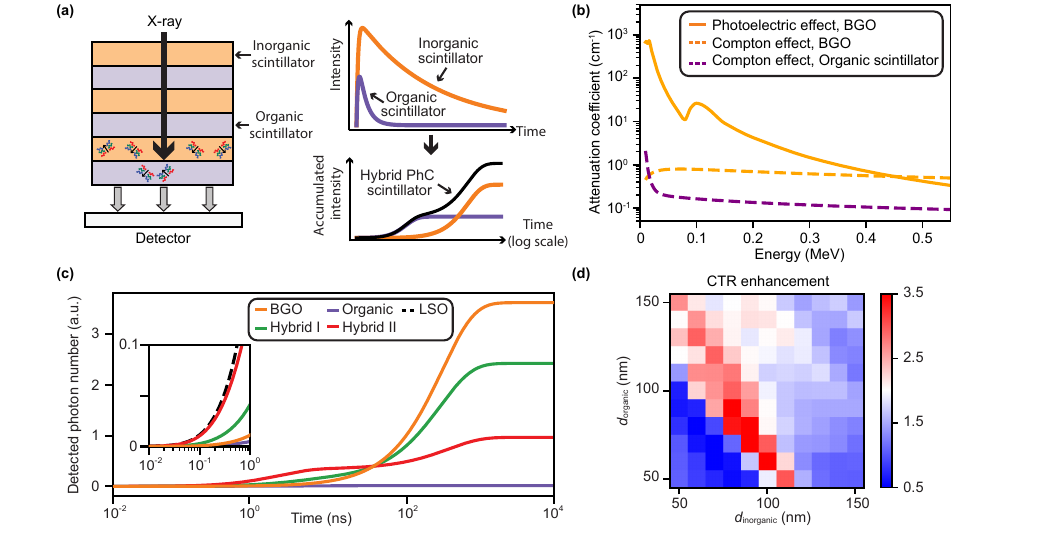}
    \vspace*{-4mm}
    \caption{\textbf{Physical mechanism of hybrid PhC scintillators.}  (\textbf{a}) Concept of hybrid PhC scintillators. Integrating organic and inorganic scintillators allows faster and brighter scintillation. (\textbf{b}) Attenuation coefficient of BGO and organic scintillator as a function of incident ionizing radiation energy. (\textbf{c}) Number of detected photons as a function of arrival time for different scintillators. Hybrid I has  $d_{\textrm{inorganic}}=\SI{150}{\nano m}, d_{\textrm{organic}}=\SI{50}{\nano m}, $ and hybrid II has $d_{\textrm{inorganic}}=\SI{70}{\nano m}, d_{\textrm{organic}}=\SI{130}{\nano m}$. Except for LSO having $\sim$\SI{9}{\micro m} to give equal X-ray attenuation length as hybrid II, all structures are \SI{20}{\micro m} thick. (\textbf{d}) Enhancement of CTR as a function of $d_{\textrm{organic,inorganic}}$. The enhancement is defined by comparing the CTR of a hybrid PhC structure consisting of 100 alternating layers of organic and inorganic scintillators to that of a bulk LSO of equivalent thickness. Number of early arriving photons $N_{\textrm{early}}$ are defined as the number of photons arriving within the first nanosecond.}
    \label{fig:hybrid}
\end{figure*}

Fig.~\ref{fig:ITO_plastic}(c) shows the dependence of the number of scintillating photons generated within the structure on the thickness of each stopping layer. By alternating ITO and organic scintillator layers while fixing the total thickness of each material at  \SI{5}{\micro\meter} (\SI{10}{\micro\meter} in total), the overall X-ray attenuation and the conversion efficiency from X-rays to high-energy electrons remains unchanged. The total number of layers increases as $2^{N}$ while the individual ITO and organic scintillator layer thickness ($d_{\textrm{ITO,scint}}$) decreases exponentially to maintain the same total thickness. We utilize the Geant4 software~\cite{agostinelli2003geant4} to perform Monte Carlo simulation that calculates the spatial distribution of scintillating photons within the structure. Each curve maintains a plateau until $d_{\textrm{ITO,scint}}$ reaches 1\% of the X-ray attenuation length of ITO at the respective X-ray energy. In other words, $d_{\textrm{ITO}}$ should be significantly smaller than the X-ray attenuation length of ITO to maximize the number of photoelectrons that escape ITO and contribute to scintillation. For many scintillators emitting visible light, the thickness of each layer should remain on the order of few hundred nanometers to remain subwavelength, which is much smaller than the X-ray attenuation length of ITO as shown in Fig.~\ref{fig:ITO_plastic}(a). Therefore, PhC scintillators can naturally maximize the extraction efficiency of photoelectrons from the stopping layer. 

We define three enhancement metrics and show how these metrics change as a function of $d_{\textrm{ITO,scint}}$. The enhancement in the number of scintillation event inside the structure $n$ reflects the increase in the number of photoelectrons in ITO layers that contribute to scintillation. The decay rate enhancement $\Gamma$ is the spectrally and spatially averaged Purcell factor (details included in the Methods section). The total light output enhancement $\eta$ combines $n, \Gamma$ and the change in the outcoupling efficiency between the PhC and bulk structures. As shown in Fig.~\ref{fig:ITO_plastic}(d), a strong photoelectric absorption and a high refractive index of ITO enables $n,\Gamma,\eta$ to reach up to 570, 1.4, and 700 respectively.  

So far, we have explored how 1D PhC scintillators can enhance scintillation light output and decay rate simultaneously. Our PhC scintillator is suitable for certain applications where both high scintillation yield and short decay time are necessary. One such application is TOF-PET imaging, which would be discussed in the next section.

\subsection*{Hybrid scintillators}

In TOF-PET, the timing difference between two \SI{511}{\kilo\eV} annihilation photons is measured with an array of scintillator detectors to determine the spatial position of the annihilation event~\cite{vandenberghe2016recent}. The spatial resolution is determined by the time resolution of the detector, which is also referred to as the coincidence time resolution (CTR). The CTR $\tau_{\textrm{CTR}}$ estimated from early photon-time density depends on the number of early arriving photons $N_{\textrm{early}}$ and the scintillation decay time $\tau_d$: $\tau_{\textrm{CTR}} \propto \sqrt{\frac{\tau_d}{N_{\textrm{early}}}}$~(Ref.~\citenum{pagano2024nanocrystalline,gundacker2019high,lecoq2010factors,regis2025gamma}.) Therefore, scintillators with both high scintillation output and fast decay time are critical for TOF-PET applications. Commercial TOF-PET scanners use inorganic scintillators such as lutetium oxyorthosilicate (LSO), which has higher scintillation light output that can outweigh its slower decay rate compared to organic scintillators~\cite{schaart2021physics}. 

We propose a ``hybrid'' scintillator that integrates inorganic and organic scintillator layers into a PhC structure (Fig.~\ref{fig:hybrid}(a)), which further improves the overall scintillation light output compared to the PhC scintillator with a non-scintillating stopping layer described in the previous section. Similar to non-scintillating stopping layers, inorganic scintillators have higher X-ray absorption and higher refractive indices than organic scintillators. In addition, high-energy electrons that do not reach the organic scintillator layers also contribute to scintillation when non-scintillating stopping layers are replaced by inorganic scintillators, further increasing the overall scintillation light output. Therefore, hybrid PhC scintillators can leverage the advantage of both scintillator types: the fast decay time of the organic scintillator and the high scintillation yield of the inorganic scintillator -- a combination that also motivates the ``metascintillator'' concept~\cite{shultzman2024towards}. Beyond this combination, our hybrid PhC scintillator additionally leverages Purcell-enhanced emission enabled by nanoscale stopping layers and the photonic crystal structure. 

To demonstrate how hybrid PhC scintillators can be applied for TOF-PET applications, we design hybrid PhC scintillators integrating organic scintillator with bismuth germanate (BGO). BGO is suited as a stopping layer due to its high refractive index ($n=2.15$ at its peak emission wavelength)~\cite{brunner2017bgo} and superior stopping power at \SI{511}{\kilo eV} compared to organic scintillators, as illustrated in Fig.~\ref{fig:hybrid}(b). Although the ratio of the X-ray attenuation power between BGO and the organic scintillator decreases as the incident ionizing radiation energy increases, the attenuation power of BGO remains 10 times higher than that of the organic scintillator at \SI{511}{\kilo\eV}. 

Figure~\ref{fig:hybrid}(c) shows the number of outcoupled scintillating photons as a function of arrival time for several structures. While BGO has a higher stopping power at \SI{511}{\kilo\eV} than LSO, its relatively low scintillation yield and slow decay time ($\sim$\SI{300}{\nano\second})~\cite{BGO} makes it unsuitable for the TOF-PET application (orange line in Fig.~\ref{fig:hybrid}(c)). TOF-PET scanners using bulk organic scintillators have been proposed~\cite{moskal2012tof}, which are still limited by low scintillation yield (purple line). Combining these two scintillators into a hybrid PhC scintillator enhances the number of outcoupled photons generated by the organic scintillator through Purcell enhancement and the additional high-energy electrons generated by BGO (red line). By adjusting the structural parameters (i.e., the thickness of each type of scintillator) while keeping the total thickness constant, one can balance the number of fast-arriving photons and the total light output (green line). Compared to LSO with the same X-ray attenuation length as the hybrid PhC scintillator (black dashed line in the inset panel), the hybrid structure generates a similar number of early arriving photons with a faster decay time.

Fig.~\ref{fig:hybrid}(d) illustrates the CTR enhancement achieved relative to a bulk LSO of equivalent geometrical thickness. Although the total light output of the hybrid PhC scintillator is lower than that of the bulk LSO, the hybrid structure achieves up to 20 times faster emission while generating approximately 50\% of the early arriving scintillating photons compared to the bulk LSO. This combination leads to a CTR enhancement up to a factor of 3.5.

\section{Discussion}
The presented work demonstrates that both the total scintillation light output and the emission rate of the scintillator can be simultaneously enhanced by integrating a stopping layer with an organic scintillator to form a 1D PhC structure. The proposed design supports both efficient photoelectric absorption and Purcell enhancement.

While the PhC scintillator exhibits a significant enhancement in total scintillation light output compared to a bulk organic scintillator, its overall brightness remains lower than that of a bulk inorganic scintillator of equivalent thickness as shown in Fig.~\ref{fig:hybrid}(c).  In particular, for PhC scintillators with non-scintillating stopping layers, high-energy electrons generated within the stopping layer have to escape into the organic scintillator layers to produce scintillating photons, which inevitably introduces the loss in the number of high-energy electrons that contribute to scintillation. 

Although we kept the total thickness to be the same for most of the cases when comparing two different types of scintillators, other metrics including weight and cost could be considered when evaluating the overall performance of the scintillator. In CT scanners, the total mass of the scintillator and its distribution are critical for maintaining mechanical balance, as the detector array has to rotate to acquire tomographic images at different angles~\cite{cramer2018stationary}. For PET scanners, organic scintillators are more cost efficient than the inorganic scintillators due to their low melting point and the abundance of their constituent materials. In our PhC scintillators, the organic scintillator has approximately 7 times lower density than LSO. For the type II hybrid PhC scintillator in Fig.~\ref{fig:hybrid}(c), a bulk LSO of equal weight would be only 36\% as thick as the hybrid PhC scintillator. 

Before concluding, we also discuss potential approaches for fabricating the proposed PhC scintillators. Organic scintillators have low melting points, which provides additional flexibility in fabrication~\cite{knoll2010radiation}. The first approach involves infiltrating liquid organic scintillator into a photonic crystal mold fabricated from an inorganic material. Such molds can be made by direct laser writing~\cite{rodenas2019three} or reactive-ion etching~\cite{takahashi2009direct}. The second approach is to fabricate the PhC scintillator directly without infiltration. Studies on fabricating different 1D inorganic-organic hybrid nanophotonic structures including distributed Bragg reflectors (DBRs)~\cite{shi2015conductive} and PhC slabs~\cite{wang2010organic} can be adapted to fabricate our PhC scintillators. Recently, fabrication of a 1D heterostructure nanoscintillator consisting of organic scintillator and titanium dioxide (\ce{TiO2}) was demonstrated using spin-coating \cite{be2025heterostructure}. Scaling hybrid PhC scintillators to the size required for PET scanners can be more challenging, as typical PET scanners require sub-centimeter-thick scintillators to efficiently absorb the incident radiation~\cite{poon2012optimal}. One possible alternative is to encapsulate the PhC scintillator within bulk stopping layers~\cite{shultzman2024towards}.

\section{Methods}

\subsection{X-ray simulation}
Geant4 is a Monte Carlo software that models scintillation by calculating the energy deposition inside the material based on high-energy particle and matter interaction. In Geant4, the energy deposition is determined by the material's chemical composition and density, and the scintillation behavior is determined by its scintillation yield and the emission spectrum. Material parameters can be found in Ref.~\citenum{BGO}, ~\citenum{chen2020transparent,lohner2014optical,EJ296,kumar2011thickness,kapusta1999comparison,melcher1992cerium}. ITO is modeled as the mixture of 90 wt\% $\textrm{In}_2\textrm{O}_3$ and 10 wt\% $\textrm{Sn}\textrm{O}_2$. We assume visible light absorption is negligible for all the materials we use for PhC scintillator. After we define the material properties of scintillators and stopping layers, we design a PhC scintillator, inject either hard X-rays (Fig.~\ref{fig:ITO_plastic}) or $\gamma$-rays (Fig.~\ref{fig:hybrid}), and simulate the scintillation behavior.

\subsection{Nanophotonic simulation}

The current Geant4 framework does not account for the modification of spontaneous emission due to the change in the photonic environment. Therefore, we evaluate the emission rate enhancement and outcoupling efficiency through a separate calculation, following the method described in Ref.~\citenum{kurman2020photonic}, \citenum{shultzman2023enhanced}. 

The spatially and angularly dependent Purcell factor of the finite 1D PhC structure is calculated using the effective Fresnel coefficients from multilayered media~\cite{wasey2000efficiency}. The effective emission rate is subsequently determined by integrating the Purcell factor along spatial, angular, and spectral domains. For the spatial and spectral averaging, the integration is weighted by the spatial distribution of emitters within the structure and the emission spectrum of the scintillator. This approach ensures that the calculated Purcell enhancement represents the spatial distribution of emitters within the 1D PhC.

\section{Data and code availability statement}

All the data and codes that are used within this paper are available from the corresponding authors upon request. Correspondence and requests should be addressed to S.~C. (seouc130@mit.edu).

\bibliographystyle{unsrt}

\bibliography{bibliography.bib}

\section{Authors contributions}
S.~C., S.~V., C.~R.-C., and M.~S. conceived the original idea. S.~C. developed the framework for nanophotonic scintillators with the help from A.~S.; S.~C. acquired and analyzed the simulation data with contributions from S.~V., and C.~R.-C.; M.~S. supervised the project. The manuscript was written by S.~C. with inputs from all authors.

\section{Competing interests}

The authors declare no competing interests.

\section{Acknowledgements}

We thank S. Min, S. Pajovic, J. Chen, and W. Michaels for stimulating discussion. S.~C. acknowledges support from Korea Foundation for Advanced Studies Overseas PhD Scholarship. C.~R.-C. is supported by a Stanford Science Fellowship. The authors acknowledge the MIT SuperCloud and Lincoln Laboratory Supercomputing Center for providing computation resources supporting this project. This material is based upon work also supported in part by the U.S. Army Research Office through the Institute for Soldier Nanotechnologies at MIT, under Collaborative Agreement Number W911NF-23-2-0121. During the preparation of the manuscript, the authors used Claude, developed by Anthropic to improve the readability and language of the work. The authors reviewed and edited the AI-assisted text and take full responsibility for the content of the publication.

\end{document}